\documentclass[10pt,final,journal,letterpaper,twoside,twocolumn]{IEEEtran}
\usepackage{amsfonts}
\usepackage{amssymb}
\usepackage{amsmath}
\usepackage{dsfont}
\usepackage{graphicx,subfigure}
\usepackage{enumerate}
\usepackage{cite}
\usepackage{bigstrut}
\usepackage{float}
\usepackage{balance}
\usepackage{color}
\usepackage{lipsum}
\usepackage{psfrag}
\usepackage{algorithm}
\usepackage[noend]{algpseudocode}
\usepackage[export]{adjustbox}

\begin{document}
%\IEEEoverridecommandlockouts

\title{Wireless Federated Learning (WFL) for 6G Networks - Part I: Research Challenges and Future Trends}
\author{Pavlos S. Bouzinis, \IEEEmembership{Student Member,~IEEE}, Panagiotis D. Diamantoulakis,  \IEEEmembership{Senior Member,~IEEE}, and George K. Karagiannidis,  \IEEEmembership{Fellow,~IEEE}
\thanks{P. S. Bouzinis, P. Diamantoulakis,  and G. Karagiannidis are with Wireless Communication and Information Processing  Group (WCIP), Department
of Electrical and Computer Engineering, Aristotle University of Thessaloniki, Greece, E-mails: \{mpouzinis, padiamian, geokarag\}@auth.gr}
}
\maketitle

\begin{abstract}
 Conventional machine learning techniques are conducted in a centralized manner. Recently, the massive volume of generated wireless data, the privacy concerns and the increasing computing capabilities of wireless end-devices have led to the emergence of a promising decentralized solution, termed as \textit{Wireless Federated Learning (WFL}). In this first of the two parts paper, we present the application of WFL in the sixth generation of wireless networks (6G), which  is envisioned to be an integrated communication and computing platform. After analyzing the key concepts of WFL, we discuss the core challenges  of WFL imposed by the wireless (\textbf{or mobile communication}) environment. Finally, we shed light to the future directions of WFL, aiming to compose a constructive integration of FL into the future wireless networks.
\end{abstract}
\vspace{-0.2in}
\vspace{-0.2in}
\section{Introduction}
\IEEEPARstart{S}{ixth} generation of wireless networks (6G) is envisioned to be an integrated communication and computing platform, with the capability to serve a vast amount of heterogeneous internet-of-things (IoT) applications, e..g, autonomous vehicles, augmented and virtual reality, smart grids, intelligent industry, smart farming, etc. To this direction, the main pillar is the twofold use of artificial intelligence, both as a means to  efficiently orchestrate the wireless networks and as the core of the operation of most applications. This separation steers the research in different  directions, namely: i) the optimization of wireless networks performance by using machine learning (ML) techniques and ii) the enhancement of data-driven applications that are based on ML, by the joint design and optimization of communication and computing networks \cite{6G}. This work mainly focuses on the second direction, although the use of techniques from the first direction is also considered.

Nowadays, the standard ML techniques are based on a centralized concept, where the data are uploaded and processed on a single entity, e.g., a central server. However, the strict latency requirements and the data privacy assurance, renders the centralized configurations impractical for forthcoming applications, such as smart grids, autonomous vehicles, and augmented reality. Hence, the combination of the aforementioned limitations with the growing computational capabilities of devices, paves the way towards implementing distributed frameworks for the construction of learning models. In the decentralized solutions, devices collaboratively train a model by leveraging their local computational resources. Among the decentralized approaches, federated learning (FL) has been proposed as a promising solution for protecting the data privacy and meeting the low-latency demands \cite{konevcny2016}, \cite{mcmahan2017}.
\par
The prominent feature of FL is the retention of the training dataset in the source of generation i.e., the device.  More specifically, each learner performs the model training through its local dataset individually, and forwards only the training parameters to the central server, instead of sending the overall raw data. In this manner, the central unit has no explicit access to privacy-sensitive data. Following that, the server is aggregating the received parameters, updates the global model and finally broadcasts it to the learners, while the considered process is repeated until convergence of the global model. 

The principal advantages that FL is capable of providing are discussed below.

%\vspace{-0.1in}
\begin{itemize}
	\item \textit{Privacy}: As mentioned previously, users do not share their raw data with the server or any of the residual participants. Therefore, the privacy-preserving mechanism constitutes an inherent characteristic of FL. 
	\item \textit{Very low Latency}: Since no raw data are sent to the cloud, the amount of information transmitted into the network is reduced, which also decreases the communication cost. Furthermore, decisions and model training can be executed locally on the end devices, instead of being sent to the server, leading to decreased latency.
	\item \textit{System Heterogeneity}: The devices participating in the learning process, might present heterogeneity in terms of computational, communication resources, and data heterogeneity, which deals with non-independent and identical distribution (non-i.i.d.) of data among users. FL has the potential to tackle with the former issues. 
\end{itemize}
Next, the general wireless federated learning (WFL) reference architecture is presented, which is the basis of the considered core WFL applications, as well as the analysis presented in the second part of this work.
\begin{figure}[t!]
\vspace{-0.1in}
	\centering
	\includegraphics[width=0.75\linewidth]{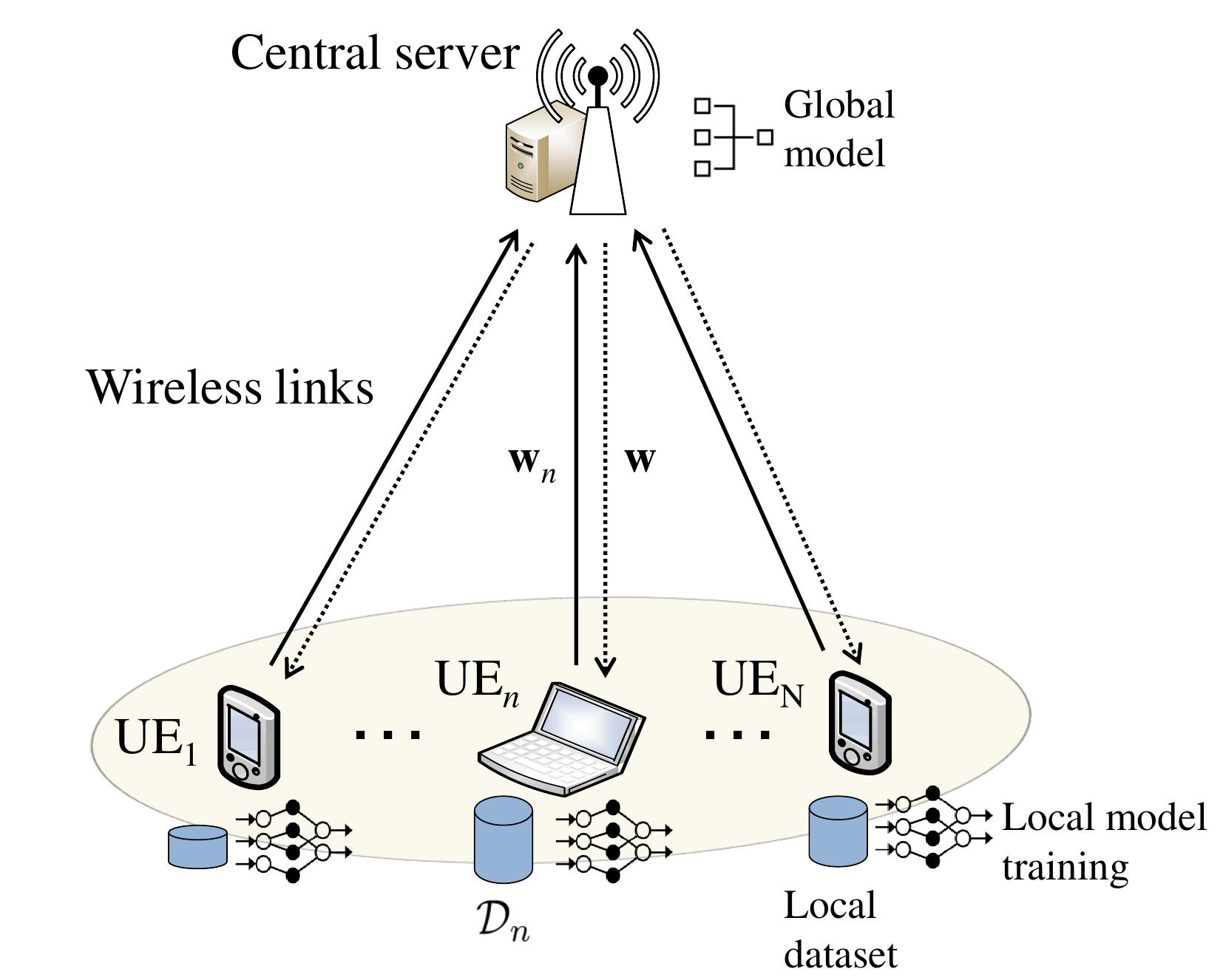}
	\vspace{-0.2in}
	\caption{Reference architecture of WFL.}
	\label{sm}
	\vspace{-0.2in}
\end{figure}
\vspace{-0.1in}
\subsection{Wireless federated learning reference architecture}
As previously mentioned, FL refers to training a shared model in a distributed manner, by exploiting the collected data of the  mobile devices without those being intervened by the server. Hence, each device contributes to the construction of the model by performing local training on its dataset, while the server's role is to aggregate, update, and redistribute the updated model back to the users. As a result, each user benefits from the local datasets of the residual participants, with the aid of a central server, while the data privacy is preserved.
\par
We consider a WFL network, which consists of $N$ users indexed as $n \in \mathcal{N}=\{1,2,...,N\}$ and a BS-server. Each user $n$ has a local dataset $\mathcal{D}_n=\{\boldsymbol{x}_{n,k},y_{n,k}\}^{D_n}_{k=1}$,  where $D_n=\vert \mathcal{D}_n \vert$ are the data samples, $\boldsymbol{x}_{n,k}\in \mathbb{R}^d$ is the $k$-th input data vector of user $n$, while $y_{n,k}$ is the corresponding output. The whole dataset is denoted as $\mathcal{D}=\underset{n \in \mathcal{N}}{\cup }\mathcal{D}_n$, while the size of all training data is given as $D=\sum_{n=1}^{N}D_n$. Following that,
the local loss function on the data set $D_n$,  defined as
\begin{equation}
F_n(\boldsymbol{w}) \triangleq \frac{1}{D_n}\sum_{k \in \mathcal{D}_n}f(\boldsymbol{w},\boldsymbol{x}_{n,k},y_{n,k}), \quad \forall n \in \mathcal{N},
\end{equation}
where $f(\boldsymbol{w},\boldsymbol{x}_{n,k},y_{n,k})$ captures the error of the model parameter $\boldsymbol{w}$ for the input-output pair $\{\boldsymbol{x}_{n,k},y_{n,k}\}$. The aim of the training process is to find the global model parameter $\boldsymbol{w}$ which minimizes the loss function on the whole data set, which is given by $J(\boldsymbol{w})=\frac{1}{D}\sum_{n=1}^{N}D_nF_n(\boldsymbol{w})$,
%\begin{equation}
%J(\boldsymbol{w})=\frac{1}{D}\sum_{n=1}^{N}D_nF_n(\boldsymbol{w}),
%\end{equation}
i.e., to find $\boldsymbol{w}^*= \underset{\boldsymbol{w}}{\mathrm{argmin}}\,J(\boldsymbol{w}).$
\par
The whole training process is divided in an arbitrary number of communication rounds, denoted by $i$. Thus, the $i$-th round is described by the following steps \cite{mcmahan2017}:
\begin{enumerate}[i)]
	\item The BS broadcasts wirelessly the global parameter $\boldsymbol{w}^i$ to all users during the considered round.
	\item After receiving the global model parameter, each user $n \in \mathcal{N}$, train its local model by applying a few steps of the gradient descent method, i.e., $\boldsymbol{w}^{i+1}_n=\boldsymbol{w}^i-\eta \nabla F_n(\boldsymbol{w}^i)$, where $\eta$ is the learning rate, and then uploads the local parameter $\boldsymbol{w}^{i+1}_n$ to the server. It is noted that, alternative methods could be also employed for the local training, such as stochastic gradient descent (SGD).
	\item After receiving all the local parameters, the server aggregates them, in order to update the global model parameter, by applying $\boldsymbol{w}^{i+1}=\frac{1}{D}\sum_{n \in  \mathcal{N} }D_n\boldsymbol{w}^{i+1}_n$.
\end{enumerate}
The above process is repeated until a required global accuracy is achieved. Moreover, during the first round the server initializes  $\boldsymbol{w}^0$. Fig. \ref{sm}, illustrates the FL architecture in a wireless network.
\subsection{Applications of wireless federated learning}
%\vspace{-0.1in}
The range of WFL applications is quite large and not fully explored yet. Next, three core applications in the era of 6G are discussed, emphasizing on their particularities and requirements.
\subsubsection{Smart grids}
 Smart grids can be seen as the superposition of electricity and communication networks, which enables the two-way flow of power and data, facilitating the active participation of all users in the energy management, the precise  prediction of energy consumption, the avoidance of security risks, the self-healing procedure, etc. This approach leads to the generation and the requirement of processing of an enormous amount of data, which might be difficult or even impossible to be stored and processed centrally \cite{SGbigdata}. Also, despite the important benefits of the electricity networks intellification, the exchange of information between different entities and the processing of data at the cloud, exposes the smart grids to potential security and privacy risks. To overcome these challenges, the iterative local processing of data by the  smart meters and aggregators at the edge and the global exploitation of the corresponding output in a collaborative manner via  WFL can be particularly useful \cite{electricity}. 
The use of WFL in a power grid mobile edge computing environment has been considered in \cite{smartgrid}, where a dynamic client selection optimization framework was presented that takes into account the time-varying link reliability. Also, in \cite{electricvehicles} a WFL approach was presented that facilitates the energy demand prediction for electric vehicles.
  \subsubsection{Unmanned mobility}  Autonomous vehicles are an emerging application which is envisioned to be realized with the aid of 6G wireless networks. In order to support coordination among vehicles and satisfy the requirements of unmanned vehicular networks, machine learning-based techniques constitute a significant tool. Relevant applications are the collaborative autonomous driving, collision avoidance systems, visual object detection and traffic congestion control. Usually, the model training is performed at a central cloud in an off-line manner. However, such approaches cannot adapt to the dynamic system changes. Therefore, WFL could alleviate this burden, as a highly adaptive technique which monitors the environmental changes in real-time. Furthermore, each vehicle could benefit from the rest vehicles' observations, in a collaborative manner, leading to increased environmental knowledge \cite{du2020}. At last, FL has the potential to reduce the data traffic, which is vital for latency-critical applications such as autonomous vehicles.
  \subsubsection{Augmented reality (AR)} In the past years, AR technology has received significant attention. AR provides an interactive experience to the user, by combining virtual contents with the real world \cite{chen}. Traditionally, AR models are trained in a centralized manner. However, the latency-sensitive AR applications impose new challenges, while the centralized machine learning approach becomes non applicable. Therefore, FL is capable of providing low-latency for object detection tasks and classification problems. Also, in \cite{chen}, the FL concept has been combined with mobile edge computing (MEC), in order to exploit the computing capabilities of edge nodes and reduce the computational power consumption at the end device during the WFL process.

\section{Challenges}
The implementation of FL in wireless networks faces several distinctive challenges. In this section, some core challenges that come along with the realization of the FL concept are discussed.
\subsection{Resource allocation and participants selection}
In a wireless FL environment, one of the major issues that needs to be addressed is the management of the available computation and wireless resources that users share. Optimal resource allocation strategies can lead to decreased latency per round and subsequently to fast convergence of the global FL model. Firstly, the scarce spectrum resources ought to be efficiently allocated among users. Furthermore, since many devices are energy-limited, the power control and the computation energy consumption of the devices have to also be considered. Therefore, in order to meet the strict latency, energy and training efficiency requirements, the bandwidth allocation, the transmission power control and the devices' CPU frequency clock speed adjustment issues, need to be jointly orchestrated. It is worth noting, that the resource allocation problem is highly related with the multiple access protocol selection, which we will discuss later on this work. For instance, authors in \cite{yang2020} minimized the total energy consumption of all users during an FL task, under latency constraints, by optimally managing the available computation and communication resources.
\par
Besides resource management, the number of participating users in the FL task has to be tactfully selected. During a communication round, the server waits until all participants terminate the update and upload of their locally trained models. Therefore, the delay of a round is determined by the slowest device. As a result, devices with limited computation capabilities or poor wireless channel conditions, termed as stragglers, are responsible for the occurrence of long delays and can negatively affect the convergence speed. In addition to this, the limited number of resource blocks may not be adequate to support an increased number of devices. Furthermore, the non-i.i.d. level of the overall dataset among users, impacts the number of selected clients. As a consequence, only a subset of the eager-to-participate users may be scheduled for participation. Thus, device scheduling policies become crucial for satisfying the underlying latency constraints, accelerating the model convergence and improving the model performance.
%\vspace{-0.1in}
\subsection{Tradeoff between latency per round and number of total rounds}
In a FL task, one of the objective goals is to minimize the convergence time, in order to achieve a certain global accuracy. The convergence time is a function of the total number of rounds and the latency per round, which is subject to both computation and communication delay. As a matter of fact, during the local model training process, the number of local iterations that each device performs has to be wisely selected. Increased number of local iterations may lead to decreased number of required rounds, in the expense of energy consumption and larger latency per round. On the other hound, the execution of few local iterations are energy-saving and achieve smaller latency per round, however an increased number of total communication rounds may be enforced, in order to achieve the required global accuracy. Moreover, the considered tradeoff is also present during the participant selection procedure. By scheduling a large amount of users for participation, an increased latency per round is expected owing this to the straggler effect and the reduced bandwidth allocation. On the contrary, by selecting a few users to participate, it is more likely that the latency per round is decreased. However, in such case, the convergence speed and the model accuracy might be negatively affected, since a limited number of scheduled users contributes with a smaller dataset throughout the training process.
\subsection{Tradeoff between model performance and convergence speed}
Apart from convergence rate, global model accuracy is of paramount importance, since it is the primary goal of the training process. The inherent unreliability of wireless links can impact the quality of the WFL performance. Motivated by such considerations, authors in \cite{chen2020} investigated the effect of packet transmission errors, aiming to improve the FL performance, by jointly optimizing the computation and radio resources, as well as the user selection. Furthermore, in \cite{shi2020}, the global loss function of the FL model is minimized subject to convergence time constraints. Emphasis is given to the bandwidth allocation and the user scheduling policy. Moreover, the non-i.d.d. level among the local datasets has be shown to significantly impact the user selection decisions, for a timely-efficient model performance improvement. Finally, the number of local updates that each device executes can affect the global model performance. By performing a few local iterations, a decreased global performance may occur. Reversely, the local over-optimization, i.e., large number of local updates, could lead to divergence and deterioration of the global model accuracy \cite{mcmahan2017}. As a matter of fact, a balanced number of local updates should be selected, while the tradeoff between convergence speed and model accuracy should be also considered.
\subsection{Privacy and security}
Guaranteeing the privacy of the local datasets is a fundamental driving factor for implementing WFL. Although the participants do not share their raw data, sensitive information may still be revealed to malicious third-partys or the central server e.g., with gradient leakage attacks to steal the devices' local data. The privacy in FL systems can be classified into the following two categories: global privacy and local privacy. Global privacy requires that no third party can access the global model during each communication round, while local privacy requires that the updates of the model are also private to the server. Differential Privacy (DP) techniques \cite{DP}, have been proposed to protect gradient information, while they can be used for ensuring both global and local privacy. DP is based on the addition of artificial noise in the training parameters by using a differential privacy-preserving randomized mechanism. Although DP can enhance privacy, it may sacrifice the model performance.
\par
Moreover, FL faces security issues, such as data and model poisoning attacks. Malicious participant can send incorrect information or false models in order to undermine the training efficiency and degrade the global model performance. Thus, protection mechanisms should be constructed, which aim to detect abnormal user behavior and finally prevent malicious users from participating into the training process.
\subsection{Dynamic wireless environment}
Wireless links are unreliable and can often vary through time. These dynamic changes can affect users willingness regarding participation throughout the whole training process. Specifically, active devices may be obligated to drop out the training process in an arbitrary time instant, due to connectivity issues related with bad channel conditions or energy-intensive tasks. To circumvent this challenge, adaptive methods should be adopted, which dynamically orchestrate the WFL network during the training procedure. For instance, the number of scheduled clients may differ among communication rounds. Also, another approach is to aim at controlling the channel uncertainty by using cooperative communication technologies or intelligent reflective surfaces \cite{IRS}. Moreover, since many devices are unreliable of successfully completing a task, asynchronous communication schemes may be considered, where the server does not necessarily wait for all participants to finish the parameter transmission \cite{xie}.

\section{Future Research Directions}
%This section shed lights on the future topics concerning the implementation of federated learning in wireless networks.
\subsection{Advanced multiple access for WFL}
During the model transmission phase in each communication round, all devices upload their locally trained results to the central server. Thus, the efficient integration of WFL in 6G depends on the utilized multiple access scheme.  In the recent literature, mostly orthogonal multiple access (OMA) schemes are selected for the considered uplink transmission, such as frequency division multiple access (FDMA) and time division multiple access (TDMA). In the former scheme, each user occupies a sub-channel from the available bandwidth, while in TDMA, users are  transmitting  their messages in non-overlapping time slots by utilizing the whole available bandwidth. 
\par
In the last years, non-orthogonal multiple access (NOMA) has drawn considerable attention, as a spectral-efficient multiple access technique \cite{ding2017}. Apart from spectral efficiency, NOMA is capable of increasing the number of  served devices and also providing fairness among users. Due to the aforementioned capabilities, NOMA has the potential to reduce the communication cost during the WFL task. In this direction, a NOMA paradigm is investigated in the second of the two parts, where the  \textit{Compute-then-Transmit NOMA (CT-NOMA)} protocol is introduced and optimized. According to  CT-NOMA  the users terminate concurrently the local model training and then simultaneously transmit the trained parameters to the central server \cite{part2}.
 It should be  highlighted that hybrid NOMA/OMA configurations are also worth of investigation, aiming to further reduce the latency and meet the WFL demands, by capitalizing on both the orthogonal and non-orthogonal aspects of multiple access and their underlying advantages. An example of the later considerations, could be the scheduling of few devices in the same orthogonal resource block, combined with the utilization of NOMA.
\subsection{WFL over fog radio access networks}
In order to satisfy the versatile requirements of 6G wireless networks, fog radio access networks (FRANs) have been proposed as a promising network architecture to provide low latency and massive connectivity. The edge nodes provided by the FRANs, which are empowered with powerful computation capabilities, could be an effective tool for assisting the training process during WFL. Firstly, FRANs can decongest the local devices from computationally-intensive tasks. Secondly, the provision of massive connectivity can lead to reduced communication cost, while it can also scale up the number of participating devices. Therefore, the FRAN is visualized as an intermediate layer between the participants and the cloud server, aiming to improve the training efficiency, by contributing with computational power and dealing with device density and connectivity issues.
\par
A promising technique which leverages the aforementioned architecture is the Hierarchical Federated Learning (HFL) \cite{abad2020}. According to HFL, users update their local model and send the them to the fog nodes wirelessly. At this stage, the local models are aggregated by the fog nodes, i.e., a local averaging of the models is performed by the edge nodes. Following that, each fog node forwards the local averaged model to the central cloud via fronthaul links. Finally, the cloud server acts as a global aggregator and generates the global average model, which is reported to the users through the fog nodes, while the process is repeated until convergence. The considered multi-level configuration, could offer an efficient model exchange compared to the classical client-cloud architecture. However, the ubiquitous challenges of WFL, such as the straggler effect and device scheduling, still need to be resolved.
\subsection{Asynchronous communication}
Communication bottleneck is a significant burden for the WFL implementation, leading to increased delay. In the conventional synchronous communication protocol, the latency of each round is determined by the slowest device. Thus, the latency per round and subsequently the model convergence speed is susceptible to the straggler effect. Furthermore, in case of clients' task completion failure, the progress of the model is wasted. To mitigate those phenomena, the asynchronous FL has been proposed as a promising solution \cite{xie}. The asynchronous configuration allows participants to join the FL task in an arbitrary time instant, even if a training round is still in progress. The considered salient characteristic of asynchronous communication is representative of a practical WFL implementation, while it could further enhance the scalability of WFL. However, the bounded-delay assumption which is usually made, is unrealistic for practical FL systems. Therefore, novel asynchronous communication schemes should be developed, which also take into account users' behavior and their ability to successfully complete the assigned task. 
\subsection{Towards intelligent WFL implementation}
%\vspace{-0.1in}
It is evident that, in order to ensure an efficient FL implementation over 6G wireless networks, a plethora of performance metrics should be considered and optimized. Hence, the network management and orchestration in a wireless FL setting, constitute an overriding, yet demanding issue to tackle with. More specifically, the optimization problems corresponding to the network control in WFL systems, such as resource allocation, power control, user scheduling, etc., are usually non-convex due to the coupling of several optimization variables. Moreover, the considered problems are often of a combinatorial nature. Thus, even by applying convex transformations, the solution might come along with high computational complexity, which is a major limitation and can significantly increase the overhead. Thus, conventional optimization techniques may be impractical for the realization of WFL. To alleviate this burden, machine learning techniques could be employed, to deal with nonconvex problems and retain low levels of computational complexity.
\par
Deep Reinforcement Learning (DRL) is a auspicious machine learning technique, which could be exploited for the FL network optimization. It deals with agents who learn from the interaction with a dynamic environment. The agent aims to maximize a cumulative reward, which can refer to any figure of merit. Thus, the agent learns how to make better decisions, by observing the evolution of the considered reward. Therefore, DRL could be a powerful tool for resource management and decision making in WFL systems, aiming to solve complex problems in a near-optimal fashion and provide computationally tractable solutions.
\vspace{-0.1in}
\subsection{WFL in the next-generation internet-of-things (NGIoT)}
Due to its inherent advantages, WFL has the potential to directly contribute to most of the foundational challenges of the next-generation IoT, including reliability, energy sustainability, scalability, future-proof security and trust, privacy-by-design, etc \cite{scoping}. To this direction, several interesting trade-offs  can be investigated, such as between model accuracy and energy consumption at the mobile devices in the training process. On the other hand, one of the key challenges for the NGIoT is ``the development of IoT data sharing and monetization enabling models and technologies" \cite{scoping}, which also remains a challenge when federated learning is used. To this direction, the motivation and potential economic benefits from the participation in the training of the federated learning model, as well as the sharing of the model's output between different stakeholders is a particularly interesting topic. In this context, a promising ledger technology is the blockchain, which refers to a public and trusted ledger, operating on a peer-to-peer network without any third party being involved. Thus, WFL could benefit from blockchain in terms of security and privacy improvement, since it eliminates the need of a central server, while the participants are collaboratively building the global learning model through a consensus mechanism \cite{nguyen2021}. Finally, FL can facilitate the construction of digital twin models of IoT devices, which has also been recognized as a research priority for the NGIoT \cite{scoping, digtwoins}.
\vspace{-0.2in}
\bibliographystyle{IEEEtran}
\bibliography{bibl}
\vspace{-0.1in}
\end{document}